\documentclass[aoas,MSNbibl,nameyear,dvips]{arximspdf}

\doi{10.1214/14-AOAS728B} % kopijuoti is laisko
\volume{8}
\issue{4}
\pubyear{2014}
\firstpage{1952}
\lastpage{1955}
\docsubty{FLA}
\referstodoi{10.1214/14-AOAS728}


\begin{document}
\begin{frontmatter}

\title{Discussion of ``Spatial accessibility of pediatric primary
healthcare: Measurement and inference''}
\runtitle{Discussion}

\begin{aug}
\author[A]{\fnms{Amelia M.}~\snm{Haviland}\corref{}\ead[label=e1]{haviland@cmu.edu}}%,
% \and
\runauthor{A.~M. Haviland}
\affiliation{Carnegie Mellon University}
\address[A]{H. John. Heinz III College\\
Carnegie Mellon University\\
5000 Forbes Avenue\\
Pittsburgh, Pennsylvania 15213\\
USA\\
\printead{e1}}
\end{aug}

% HISTORY:
\received{\smonth{8} \syear{2014}}
\revised{\smonth{9} \syear{2014}}

% ABSTRACT

% KEYWORDS
\end{frontmatter}

This paper describes the infusion of many fresh statistical ideas into
the area of spatial access to healthcare, and I hope that the procedures
described are widely implemented. What is described is a large and an
impressive applied research project incorporating space-varying
coefficient models, simultaneous confidence bands and backfitting to
address otherwise potentially unstable and computationally expensive
estimation. In my opinion, there are three high-level areas of this
work that would benefit from further development. I describe these
next, followed by much briefer descriptions of some minor quibbles I
have with the paper that the authors may want to consider.

The first area where further development could be valuable is in the
``procedure developed to systematically evaluate multiple models.'' I
commend the authors in not narrowing down the space of possible models
to a single ``best'' model and instead considering a family of
acceptable models. I also appreciate that they state clear and
reasonable criteria for deeming models to be acceptable. What I find
less satisfying is that the procedure described to summarize the
multiple models deemed to be acceptable is largely qualitative.
Thus, the ability to make accurate probability statements about the
relationships between the predictors and the outcome, over the family
of acceptable models, is lost. The issues surrounding model selection
and/or how to incorporate the information from a family of useful
models into an inferential structure are highly relevant to any
decision-making that could result from statistical modeling. This issue
was highlighted in a recent National Research Council Report evaluating
the existing research regarding deterrence and the death penalty in the
U.S. [\citet{NRC2012}]. The committee for that report, which I served
upon, concluded that large model uncertainty swamped any claims of the
presence or absence of statistical significance within any particular
model. Bradley Efron's work, ``Estimation and Accuracy After Model
Selection,'' also presented at the 2014 Joint Statistical Meetings, may
be useful to consider in this context [\citet{efron2013}].

In the particular setting of this spatial accessibility analysis, the
model uncertainty issues are due to correlations among the predictors.
This source of model uncertainty makes relationships of individual
correlated predictors to outcomes of little value. A principal
components or factor analysis may be helpful to better describe what
the predictors are whose relationship with the outcome it would be
useful to estimate.

The second direction for further development stems from a quite typical
observation about optimization procedures of the sort described in
Section~2: linear optimization procedures such as the one implemented
here output quantities with no measures of uncertainty attached to
them. The authors chose to address the sensitivity of the procedure's
outputs (accessibility measures) to small variations in the constant
values incorporated into the procedure rather than incorporating the
uncertainty in those parameter estimates into the procedure such that
they are propagated through to the outputs. I appreciate the authors'
efforts in this direction, yet find it unsatisfying as there are two
things we do not know:  (1)
whether small variations from the selected constant values for the
parameters are a good
measure of the uncertainty about the parameters and (2) how the
uncertainty in many
parameters collectively impacts the procedure's output.

There appear to be several sources of uncertainty impacting the outputs
of the optimization procedure that it would be helpful to quantify;
most of these are mentioned to some degree in the paper, although not
in this context. The first source of uncertainty is that the algorithm
does not have a unique solution and thus different runs of the
procedure result in different output. Second, distance from each
patient in a census tract to a doctor is approximated using the
centroid of the census tract the patient lives in---this provides an
estimate of the actual distance that would need to be traveled. Third,
two of the bounds used in the optimization process vary at the
individual doctor level, but data on them is available only at
aggregated levels. The authors explore the sensitivity of the resulting
spatial access measurements when they simulate individual draws of
these parameters at the doctor level from the aggregate parameters.
From this they establish that in most census tracts the impacts on the
output of this uncertainty are small. Taken a step farther, this
simulation exercise would allow them to propagate the uncertainty from
this missing data through the procedure to the output. More generally,
few of the parameters for which constant values were selected are
observed or derived from meaningful thresholds and I would expect few
to be constant in reality. For instance, the maximum number of patients
that can be seen and the minimum number needed to sustain a doctor's
practice seem likely to vary based on local operating costs and wages
as well as upon the mix of reimbursement levels the doctor is receiving
from their patient mix (and presumably correlated with the proportion
of their patients with Medicaid coverage). This suggests that it would
be useful to jointly estimate these parameters with the proportion of
Medicaid patients in doctors' patient populations.
If the variability is not incorporated into the optimization procedure,
it would be useful to more systematically discuss the justification and
consequences of this decision. In some cases this may be a complete
normative argument, for example, all people should be able to have the
same travel distance limit, and in others it would involve making a
complete case for the assumptions that the variability is of a
particular modest magnitude and that considering the variability of one
parameter at a time is sufficient and then illustrating the impact of
variations of that magnitude for each parameter on the outputs of the procedure.

The third area that could benefit from additional development is that
it is unclear where the uncertainty quantified and used for making
inferences in Section~3 of the paper comes from. As described in the
paper, the optimization procedure results in measurements (with no
uncertainty in them) of spatial access for the two groups of interest
in every census tract in Georgia. This is a census with no sampling
variability. Are the authors relying on a theoretical super population
from which their data is drawn? If so, it would be useful to say so and
describe the sampling method from the super population that they are
assuming---what is sampled from that super population and how? Is
there clustering? Or is there a different source of uncertainty than a
sample from a super population?

In the remainder of this discussion I briefly describe a small set of
quibbles with or remaining questions about the methods in this paper.
While I do not question that spatial accessibility is important, I do not
find it is clear that spatial accessibility is equally important to
financial accessibility---driving a longer way or getting a ride do
not necessarily seem comparable to having the ability to get the care
paid for once you arrive. Regarding the policy simulations, there are
limitations of the fixed nature of the optimization procedure. I am
concerned that implementing a new policy, such as increasing the
mobility of those with Medicaid insurance, impacts other parameters
currently held fixed within the optimization under potential policy
changes. In the example just given, greater access to transportation
may also affect the probability that doctors farther from large
concentrations of Medicaid patients accept any Medicaid patients and
the proportion of their patient population they would allow to be
Medicaid patients. Microsimulation models, such as are implemented in
economics and other fields, may be helpful to consider, as they would
allow the parameters to jointly vary. A second quibble regarding the
policy simulations is that the authors state that they ``target
policies that are (approximately) Pareto optimal'' but focus their
discussion on the policies of reducing the probability that doctors
accept any Medicaid patients or reducing the proportion of Medicaid
patients doctors accept---neither of which can claim to be Pareto
optimal. Regarding the initial optimization procedure, it seems that
the authors are assuming all doctors provide equal quality of care for
all patients and it would be helpful for there to be some discussion of
this and how it could be relaxed. Last, I have a few remaining
questions about the methods used. Regarding Section~3.3, given that
there is no additional data regarding the distribution of the
population within each census tract, why does the Kernel Density
Estimator provide superior estimates to the simpler calculation of
dividing the population by its land mass? Providing evidence of
superior estimates could be useful. In Section~3.4, how is
``consistency'' defined? I do not understand how the Diversity Ratio is
reported to have both constant and nonconstant shapes under different
specifications and be consistent. Also in Section~3.4, what does it
mean for a predictor which has a space-varying relationship with the
outcome in all specifications to be summarized as having a
statistically significant relationship with the outcome?

% imsref loaded by akundreckaite, 2014-10-03 13:49:52
%


%suskaldyti doi

\printaddresses
\end{document}